\documentclass[12pt]{article}
\usepackage{color}
\usepackage{latexsym}
\usepackage{amsmath}
\usepackage{amssymb}
\usepackage{eufrak}
\usepackage{euscript}
\usepackage[latin1]{inputenc}
\usepackage{pstricks}
\begin{document}
\title{Primitively divergent diagrams in $\kappa$-deformed scalar field
with quartic self-interaction}
\author{M J Neves \footnote{E-mail:mariojr@if.ufrj.br} , 
C A A de Carvalho\footnote{aragao@if.ufrj.br}, 
C Farina\footnote{farina@if.ufrj.br}, M V Cougo-Pinto\footnote{marcus@if.ufrj.br}\\
{\it Instituto de F\'{\i}sica, UFRJ, CP 68528, Rio de Janeiro, RJ, 21.941-972}
}
\date{}
\maketitle

\begin{abstract}
We obtain the primitively divergent diagrams in $\kappa$-deformed scalar 
field in four-dimen\-sional spacetime with quartic self-interaction in order 
to investigate the effect of the fundamental length $q=1/(2\kappa)$ on
such diagrams. Thanks to $\kappa$-deformation, we find that the
dimensionally regularized forms of the diagrams lead to finite results in
the limit of space-time dimension four. The effect of the deformation appears 
as a displacement of the poles in the complex plane. 
\end{abstract}

\section{Introduction}

The need for a universal invariant length parameter has been 
advocated since the early days of quantum field theory, most notably
by Heisenberg, who presented at least two 
interesting arguments to add this fundamental length, say $q$, to 
the set composed of the fundamental speed $c$ and the fundamental 
action $\hbar$ \cite{Heisenberg38}. One of the arguments states that
a natural unit of mass $1/q$ is required for the determination of 
the masses of the elementary particles in any fundamental theory of matter and radiation. The absence of an explanation for the values of the masses of elementary particles is indeed an unsatisfactory feature in any such theory 
(see, {\it e.g.}, the closing remarks 
in \cite{Feynman85}). The other argument presents $1/q$ as a natural
momentum cutoff for the divergences of the theory, and derives
this cutoff from the hypothesis that, at lengths comparable to $q$,
space-time exhibits a new kind of geometry. It is well known that
the combination of gravitational and quantum effects implies indeed
the existence of such a length, specifically of order $10^{-33}$ cm, 
at the Planck scale. More recently, a fundamental minimal length
has also been advocated as a consequence of string theory (for a review,
see \cite{Witten96}). 
Whatever the motivations to have such a fundamental length scale in a quantum
field theory, we are led to investigate the new features
of such a theory to determine whether they open the possibility of unusual and
interesting physics, or give rise to unavoidable inconsistencies.  
In this article, we are interested in the role played by a minimal length in 
the regularization of a quantum field theory.

The fundamental length $q$ can be introduced in quantum field theory 
in two natural ways: as the deformation parameter of a deformation of 
space-time symmetries, as is done in the $\kappa$-deformation of 
Poincaré algebra 
\cite{LukierskiRueggNowickiTolstoy91,LukierskiNowickiRuegg92};
or as the deformation parameter of a deformation of space-time itself, 
as is done in noncommutative space-time \cite{Snyder47}.  
The $\kappa$-deformation of the Poincaré algebra
depends on a real deformation parameter $\kappa$ which in our discussion 
is related to the fundamental length $q$ by means of $\kappa=1/(2q)$
(the factor $2$ is only a matter of convenience). 
The space-time noncommutativity is implemented by promoting space-time coordinates to hermitian operators obeying relations
of the form $[x^{\mu},x^{\nu}]=i\theta^{\mu\nu}$, 
where $\theta^{\mu\nu}$ is antisymmetric, has dimension of squared length,
and is proportional to $q^2$. 

The original motivation to define noncommutative space-time was
to control the divergences which plague quantum field theories
\cite{Snyder47}, but it was found that this noncommutativity
does not necessarily eliminate all the divergences, and that it also gives 
rise to the mixing of ultraviolet and infrared divergences (see, {\it e.g.}, 
the review \cite{Szabo03}). 
In the case of the $\kappa$-deformed Poincaré algebra 
\cite{LukierskiRueggNowickiTolstoy91,LukierskiNowickiRuegg92},
it is important to investigate to what extent 
the $\kappa$ parameter can work as a natural regulator in quantum 
field theory, or at least smooth its divergences \cite{LukierskiNowickiRuegg91,LukierskiNowickiRuegg92,LukierskiRueggZakrewski95}. The $\kappa$ parameter does occur as a natural regularizing imaginary
Pauli-Villars mass parameter in a $\kappa$-deformed scalar field theory
with quartic self-interaction \cite{KosinskiLukierskiMaslanka00}. This
$\kappa$-deformed theory is defined in the $\kappa$-deformed Minkowski space,
a noncommutative space-time obtained by duality with the so called 
$\kappa$-deformed Poincaré algebra in the bicrossproduct basis
\cite{MajidRuegg94,LukierskiRueggZakrewski95}. 

Here, we want to consider
the regularization effects of the $\kappa$-deformation in
the original proposal of a quantum field theory in a commutative space-time
whose symmetries are governed by the $\kappa$-deformed Poincaré algebra 
\cite{LukierskiNowickiRuegg92} 
in the so called standard basis \cite{LukierskiRueggZakrewski95}.
We will take such a field as a scalar with quartic self-interaction, and
call it simply $\kappa$-deformed scalar field. 
We will see that the role of this $\kappa$-deformation as a possible
regulator \cite{LukierskiNowickiRuegg91,LukierskiNowickiRuegg92,LukierskiRueggZakrewski95}
is fulfilled in a rather peculiar way.

The rest of this paper is organized as follows. In section 2, we present
the basic formalism of the $\kappa$-deformed scalar field with quartic
self-interaction. Although the $\kappa$-deformation
renders the calculations much more complicated several final results
acquire a simple form when the notation of barred indices, as defined below,
is enforced. The presentation of section 2 sets the
stage for the main result of this paper, the calculation of the primitively divergent diagrams of the theory at one loop level, which is presented in section 3. Finally, section 4 is left for the concluding remarks.

\section{The $\kappa$-deformed scalar field with quartic
self-interaction}

The $\kappa$-deformed Poincaré algebra is an associative algebra which depends on a real positive deformation parameter $\kappa$ with dimension of mass, and which reduces to the Poincaré algebra in the limit $\kappa\rightarrow\infty$ \cite{LukierskiRueggNowickiTolstoy91,LukierskiNowickiRuegg92}.
This deformed algebra has the structure of the
algebraic sector of a Hopf algebra, which also has a coalgebraic
sector of no concern to us here. We consider the so called
$\kappa$-deformation in the standard basis \cite{LukierskiNowickiRuegg92}, whose defining relations of the algebraic sector are given by
\begin{eqnarray}\label{kappadeformedalgebra}
\left[P^\mu,P^\nu\right]=0\;,\;\;\;[J^i,J^j]=i\epsilon^{ijk}J^k\;,\;\;\;
[J^i,P^0]=0\;,\;\;\;[J^i,P^j]=i\epsilon^{ijk}P^k\;,\;\;\;[J^i,K^j]=i\epsilon^{ijk}K^k\;,\nonumber\\
\left[K^i,K^j\right]=-i\epsilon^{ijk}\left(J^k \cosh\frac{P^0}{\kappa}
-\frac{P^k}{4\kappa^2}{\bf P}\cdot{\bf J} \right)\;,\;\;\;
[K^i,P^j]=i\delta^{ij}\sinh\frac{P^0}{\kappa}\;,\;\;\;
[K^i,P^0]=iP^i\; ,
\end{eqnarray}
where $P^\mu$ represents the four-momentum generators, $J^i$ the rotation generators, and $K^i$ the boost generators. Hopf algebras are also known as
quantum groups or quantum algebras (although some authors reserve those
names for particular kinds of Hopf algebras) and the interested reader can
easily find an abundant literature on the subject (see, {\it e.g.}  \cite{ChaichianDemishev96,Majid95}).

The first Casimir invariant of the $\kappa$-deformed Poincaré algebra 
(\ref{kappadeformedalgebra}) is the combination of four-momentum generators
$[2\kappa\sinh(P^0/2\kappa)]^2-{\bf P}^2$, an expression that reduces to
the first Casimir invariant of the Poincaré algebra in the limit $\kappa\rightarrow\infty$, in which the deformation disappears. 
This Casimir invariant gives rise to the dispersion relation
$[2\kappa\sinh(P^0/2\kappa)]^2-{\bf P}^2=m^2$, where $m^2$ is a scalar 
labelling the irreducible representation under consideration. 
Replacing the mass parameter $\kappa$ by the length parameter $q=1/(2\kappa)$, the limit of no deformation becomes $q\rightarrow 0$, and the dispersion relation associated to the $\kappa$-deformed Poincaré algebra is given by
\begin{equation}\label{qdispersionrelation}
\left[\frac{1}{q}\sinh(qP^0)\right]^2-{\bf P}^2=m^2 \; .
\end{equation}
A $\kappa$-deformed free scalar field is a real field $\phi$ obeying the $\kappa$-deformed Klein-Gordon equation \cite{LukierskiNowickiRuegg92},
\begin{equation}\label{preqKleinGordon}
\left[\left(\frac{1}{q}\sin(q\partial_0)\right)^2-\nabla^2
+m^2\right]\phi(x)=0 \; ,
\end{equation}
which is obtained from the Casimir invariant (\ref{qdispersionrelation})
by the usual correspondence $P_0\mapsto i\partial_0$ and
${\bf P}\mapsto -i\nabla$. We denote by $\partial_q$ the $q$-differential operator introduced in reference \cite{LukierskiNowickiRuegg92},
\begin{equation}\label{partialq}
\partial_q=q^{-1}\sin(q\partial_0) \;, 
\end{equation}
in order to write equation (\ref{preqKleinGordon}) in the more compact 
form
\begin{equation}\label{pospreqKleinGordon}
\left(\partial_q^2-\nabla^2
+m^2\right)\phi(x)=0 \; .
\end{equation}
We introduce the convention that a bar over an index means that 
its range is $\{q,1,2,3\}$. With these conventions, the $\kappa$-deformed Klein-Gordon equation is recast into the simple form
\begin{equation}\label{qKleinGordon}
\left(\partial_{\bar{\mu}}\partial^{\bar{\mu}}+m^2\right)\phi=0 \; .
\end{equation}
A Lagrangian for this $\kappa$-deformed free scalar field is given by
\cite{LukierskiNowickiRuegg92}
\begin{equation}\label{L0}
{\cal L}_0=-\frac{1}{2}\phi\partial_{\bar{\mu}}\partial^{\bar{\mu}}\phi
-\frac{1}{2}m^2\phi^2 \; ,
\end{equation}
to which we want to add a self-interaction term ${\cal L}_{I}(\phi)$ to
obtain 
\begin{eqnarray}\label{L}
{\cal L}=-\frac{1}{2}\phi\partial_{\bar{\mu}}\partial^{\bar{\mu}}\phi
-\frac{1}{2}m^2\phi^2+{\cal L}_{I}(\phi) \; .
\end{eqnarray}
To proceed to the functional quantization of this theory, we postulate
the following vacuum to vacuum transition amplitude in the presence of
an external current $J$,
\begin{eqnarray} \label{Z(J)}
Z(J) = N \! \int \mathcal{D}\phi \; \exp\left\{i\int \!\! d^4x \left[-\frac{1}{2}\phi\partial_{\bar{\mu}}\partial^{\bar{\mu}}\phi-\frac{1}{2}m^2\phi^2+{\cal L}_{I}(\phi)+J(x)\phi(x)\right] \right\} \; , 
\end{eqnarray}
where $N$ is defined to render the generating functional $Z$ normalized to unity, $Z(0)=1$. As in the non-deformed case, this functional can be written as
\begin{eqnarray} \label{Z(J)Z0}
Z(J)=\frac{\exp\left\{i\!\int \! d^4z \; \mathcal{L}_{I}
\left(\frac{\delta}{i\,\delta J(z)}\right)
\right\} Z_0(J)}{\exp\left\{i\!\int \! d^4z \; \mathcal{L}_{I} \left(\frac{\delta}{i\,\delta J(z)}\right)\right\} \left.Z_0(J)\right|_{J=0}} \; ,
\end{eqnarray}
where $Z_0(J)$ is the generating functional for the free $\kappa$-deformed field, given by the expression
\begin{eqnarray}\label{Z0}
Z_0(J)=\exp\left\{-\frac{i}{2}\!\int \! d^4xd^4y J(x)\Delta(x-y)J(y)\right\}
\; ,
\end{eqnarray}
in which $\Delta(x-y)$ is the Green function for the free propagation of the
field with the prescription $m^2\mapsto m^2-i\varepsilon$, {\it i.e.}, the inverse of the $\kappa$-deformed Klein-Gordon operator given in (\ref{qKleinGordon}),
\begin{eqnarray}\label{Delta}
\Delta=\left(\partial_{\bar{\mu}}\partial^{\bar{\mu}}+m^2
-i\varepsilon\right)^{-1}\; ,
\end{eqnarray}
which has the Fourier representation
\begin{eqnarray}\label{DeltaFourier}
\Delta(x-y)=\!\!\int\!\!\frac{d^4p}{(2\pi)^4}\frac{e^{-ip.(x-y	)}}{p^{\bar{\mu}}p_{\bar{\mu}}-m^2+i\varepsilon} \; ,
\end{eqnarray}
where we have for notational convenience used the definition
\begin{eqnarray}\label{momento}
p^{\bar{\mu}}p_{\bar{\mu}}=q^{-2}\sinh^{2}(qp^0)-{\bf p}^{2} \; .
\end{eqnarray}
The $n$-point Green function of the interacting theory is given by
the usual $n$-fold functional derivative of $Z(J)$ with respect to $iJ$,
whereas the $n$-point connected Green function is given by
\begin{eqnarray}\label{Deltanconnected}
\left.\Delta_c^{(n)}(x_1, ... ,x_n)=
\frac{1}{i^{n-1}}\,\frac{\delta^{n}W(J)}{\delta J(x_1) \delta J(x_2)...\delta J(x_n)}\right|_{J=0} \; ,
\end{eqnarray}
where $W$ is defined, as in the non-deformed case, by $Z(J)=e^{iW(J)}$. 
The one particle irreducible diagrams are
also given by the Legendre transform of $W$.

Now, let us consider the self-interaction in (\ref{L}) as quartic with 
coupling constant $g$,
\begin{eqnarray}\label{gphi4}
{\cal L}_{I}(\phi)=-\frac{g}{4!}\phi^4 \; .
\end{eqnarray}
We are interested in the two-point and four-point connected Green functions at one loop level. According to the above formalism, the two-point
function is given by
\begin{eqnarray}\label{Delta2}
\Delta_c^{(2)}(x_1-x_2)=i\Delta(x_1-x_2)\!-\!\frac{g}{2}\Delta(0)\!\int \! 
d^4z \Delta(x_1-z)\Delta(z-x_2) \; ,
\end{eqnarray}
where $\Delta(0)$ is obtained from (\ref{DeltaFourier}), and the four-point
function by
\begin{eqnarray}\label{Delta4}
\Delta_c^{(4)}(x_1,x_2,x_3,x_4)=-ig\!\int d^4z\Delta(x_1-z)\Delta(x_2-z)\Delta(x_3-z)\Delta(x_4-z)+ \nonumber \\
+g^2\frac{1}{2}\!\int d^4zd^4z^{\prime}\Delta(x_1-z)\Delta(x_2-z)[\Delta(z-z^{\prime})]^2\Delta(z^{\prime}-x_3)\Delta(z^{\prime}-x_4)+\nonumber \\
+g^2\frac{1}{2}\!\int d^4zd^4z^{\prime}\Delta(x_1-z)\Delta(x_3-z)[\Delta(z-z^{\prime})]^2\Delta(z^{\prime}-x_2)\Delta(z^{\prime}-x_4)+\nonumber \\
+g^2\frac{1}{2}\!\int d^4zd^4z^{\prime}\Delta(x_1-z)\Delta(x_4-z)[\Delta(z-z^{\prime})]^2\Delta(z^{\prime}-x_3)\Delta(z^{\prime}-x_2) \; .
\end{eqnarray}
The two-point function (\ref{Delta2}) has the following Fourier representation:
\begin{eqnarray}\label{FourierDelta2}
\Delta_c^{(2)}(x_1-x_2)=i\int \frac{d^4p}{(2\pi)^4}
\frac{e^{-ip.(x_1-x_2)}}{p^{\bar{\mu}}p_{\bar{\mu}}-m^2+i\varepsilon} 
\left[1+\frac{\Sigma_1(m^2)}{p^{\bar{\mu}}p_{\bar{\mu}}-m^2+i\varepsilon}
\right]
\; ,
\end{eqnarray}
where we have defined $\Sigma_1(m^2)=ig\Delta(0)/2$ to obtain from (\ref{DeltaFourier}) the expression
\begin{eqnarray}\label{Sigma1(m2)}
\Sigma_1(m^2)=\frac{i}{2}g\int\frac{d^4p}{(2\pi)^4}
\frac{1}{p^{\bar{\mu}}p_{\bar{\mu}}-m^2+i\varepsilon} \; .
\end{eqnarray}
The four-point function (\ref{Delta4}) has the following Fourier representation:
\begin{eqnarray}\label{FourierDelta4}
\Delta_c^{(4)}(x_1,x_2,x_3,x_4)=\!\!\int\!\! \prod_{i=1}^{4}
\left(\frac{[d^4p_i/(2\pi)^{4}]\;e^{-ip_i.x_i}}{q^{-2}\sinh^2(qp_i^0)-{\bf p}_i^2-m^2+i\varepsilon}\right)
(2\pi)^{4}\delta^{(4)}(p_1+p_2+p_3+p_4)\nonumber \\ 
\left[-ig+\frac{1}{2}g^2\!\int \frac{d^4p}{(2\pi)^{4}}\frac{1}{\left[(p-s)^{\bar{\mu}}(p-s)_{\bar{\mu}}-m^2+i\varepsilon\right](p^{\bar{\mu}}p_{\bar{\mu}}-m^2+i\varepsilon)}+ \right. \nonumber \\
\left. +\frac{1}{2}g^2\!\int \frac{d^4p}{(2\pi)^{4}}\frac{1}{\left[(p-t)^{\bar{\mu}}(p-t)_{\bar{\mu}}-m^2+i\varepsilon\right](p^{\bar{\mu}}p_{\bar{\mu}}-m^2+i\varepsilon)}+ \right. \nonumber \\
\left. +\frac{1}{2}g^2\!\int \frac{d^4p}{(2\pi)^{4}}\frac{1}{\left[(p-u)^{\bar{\mu}}(p-u)_{\bar{\mu}}-m^2+i\varepsilon\right](p^{\bar{\mu}}p_{\bar{\mu}}-m^2+i\varepsilon)} \right] ,
\end{eqnarray}
where we have used the definitions
\begin{eqnarray}\label{momentos}
(p-s)^{\bar{\mu}}(p-s)_{\bar{\mu}}=\frac{1}{q^2}\sinh^{2}[q(p^0-s^0)]-({\bf p}-{\bf s})^2\, , 
\end{eqnarray}
and the ones obtained from preceding formula by replacing $s$ by $t$ or $u$, 
being $s$, $t$ and $u$ where $s=p_1+p_2$, $t=p_1+p_3$ and $u=p_1+p_4$ 
are the Mandelstam variables. 

By simple inspection, it is seen that those expressions for the Green
functions are divergent. For example, as in the non-deformed case,
the quantity $\Sigma_1(m^2)$ in (\ref{Sigma1(m2)}) is divergent. 
Therefore, $\kappa$-deformation does not render finite the Green functions
of the theory in perturbative calculations, and we are forced to resort to
the usual process of regularization and renormalization in order to define
those functions.  

\section{The one-loop renormalization with dimensional regularization}

In order to properly define two-point and four-point Green functions
for the $\kappa$-deformed field, we will use dimensional regularization
\cite{BolliniGiambiagi72,Ashmore72,tHooftVeltman72}. Actually, we will 
follow closely the original formalism of t'Hooft and Veltman  
\cite{tHooftVeltman72}. Although it is by now a subject of textbooks
(see, {\it e.g.}, \cite{Ramond90}), we will sketch its main ideas
for a simple non-deformed example in order that its comparison with
the deformed case clarifies the novel features induced by the
deformation. Let us consider the non-deformed limit of (\ref{Sigma1(m2)}),
\begin{eqnarray}\label{Sigma1(m2)q=0}
\frac{2}{ig}\left.\Sigma_1(m^2)\right|_{q=0}=\int\frac{d^4p}{(2\pi)^4}
\frac{1}{p^\mu p_\mu-m^2+i\varepsilon} \; .
\end{eqnarray}
This is not a well defined quantity ({\it i.e.}, it is not defined at all) since it is given by a divergent integral. This divergence can be viewed as a
consequence of too many space dimensions in the measure in the
numerator of the integrand for only two powers of momentum in the denominator. From this fact, we are led to consider an expression
\begin{eqnarray}\label{Sigma1(m2,omega)q=0}
\frac{2}{ig}\left.\Sigma_1(m^2,\omega)\right|_{q=0}=
(\mu^2)^{2-\omega}\int\frac{d^{2\omega}p}{(2\pi)^{2\omega}}
\frac{1}{p^\mu p_\mu-m^2+i\varepsilon} \; ,
\end{eqnarray}
in which the dimension $2\omega$ of the measure $d^{2\omega}p$ is not
restricted to $4$ or any other positive integer, but is taken to be a real
less than 2 (and $\mu$ is the mass scale parameter). To go 
back the four-dimensional space-time case,
at $\omega=2$, we further extend $\omega$ to the complex numbers, and look for an analytic continuation of (\ref{Sigma1(m2,omega)q=0}) to the right of $\Re\omega<1$, with the usual bypassing of possible singularities in the complex $\omega$-plane. The precise meaning of the measure 
$d^{2\omega}p$ in complex dimension $2\omega$ is given by the splitting
$d^{2\omega}p=dp^0\,drd\Omega_{2\omega-1}$, where $d\Omega_{2\omega-1}$
is the element of solid angle in $(2\omega-1)$-dimensional space, and 
$0\le r<\infty$ (a different splitting was used in the original formalism
\cite{tHooftVeltman72}). In this way, expression (\ref{Sigma1(m2,omega)q=0}) is precisely given by
\begin{eqnarray}\label{Sigma1(m2,omega)q=0tH-V}
\frac{2}{ig}\left.\Sigma_1(m^2,\omega)\right|_{q=0}=
\frac{(4\pi\mu^2)^{2-\omega}}
{16\pi^4\Gamma(\omega-1/2)}\pi^{3/2}
\int_{-\infty}^{\infty} dp^0\int_0^{\infty}dr^2
\frac{(r^2)^{\omega-3/2}}{(p^0)^2-r^2-m^2+i\varepsilon} \; ,
\end{eqnarray}
which is a well defined function of $\omega$ in the domain $1/2<\Re\omega<1$,
and replaces the ill defined quantity (\ref{Sigma1(m2)q=0}). Three facts are important to notice at this point. 
The first is that, properly speaking, we are not entitled to make 
$\omega=2$ in a function whose domain is restricted by $1/2<\Re\omega<1$, 
as is the case of (\ref{Sigma1(m2,omega)q=0tH-V}). The second is that 
we are entitled to search for an analytic continuation 
of (\ref{Sigma1(m2,omega)q=0tH-V}) in the neighbourhood of 
$\omega=2$ in order to investigate the behaviour of the continuation
at the physical dimension. The third is that the properties of the continued function in this neighbourhood depend on the form of the integrand in (\ref{Sigma1(m2,omega)q=0tH-V}), a non-deformed propagator in the
present case. Now, (\ref{Sigma1(m2,omega)q=0tH-V}) can indeed be analytically continued by introducing into the integrand of (\ref{Sigma1(m2,omega)q=0tH-V})
a factor of $1$ in the form 
$(\partial p^0/\partial p^0+\partial r^2/\partial r^2)/2$,
and performing appropriate partial integrations to arrive at the function
of $\omega$
\begin{eqnarray}\label{Sigma1(m2,omega)q=0prefinal}
\frac{2}{ig}\left.\Sigma_{1}(m^2,\omega)\right|_{q=0}=
\frac{(4\pi\mu^2)^{2-\omega}}
{16\pi^4\Gamma(\omega-1/2)}
\frac{\pi^{3/2}\,m^2}{\omega-1}
\int_{-\infty}^{\infty} dp^0\int_0^{\infty}dr^2
\frac{(r^2)^{\omega-3/2}}{[(p^0)^2-r^2-m^2+i\varepsilon]^2} \; ,
\end{eqnarray}
whose domain is the strip $1/2<\Re\omega<2$ punctured at the pole $\omega=1$.
Since this domain does not include a neighbourhood of the point $\omega=2$ 
of the physical dimension, we need a further analytic continuation which 
is done by the same method. Now we get the function 
\begin{eqnarray}\label{Sigma1(m2,omega)q=0final}
\frac{2}{ig}\left.\Sigma_{1}(m^2,\omega)\right|_{q=0}=
\frac{(4\pi\mu^2)^{2-\omega}}
{16\pi^4\Gamma(\omega-1/2)}
\frac{\pi^{3/2}\,2m^4}{(\omega-1)(\omega-2)}
\int_{-\infty}^{\infty}\!\!\!\! dp^0\int_0^{\infty}\!\!\!\!\!dr^2
\frac{(r^2)^{\omega-3/2}}{[(p^0)^2-r^2-m^2+i\varepsilon]^3} \; ,
\end{eqnarray}
whose domain is the strip $1/2<\Re\omega<3$ punctured at $\omega=1$
and $\omega=2$, in which the function has simple poles. In this way,
we obtain the well-known result that at four space-time dimension the 
quantity $\Sigma_1(m^2,\omega=2)$ to be absorbed into the renormalized 
mass is infinite. With this elementary example in mind, let us proceed
to the $\kappa$-deformed case. 

As we have seen, the $\kappa$-deformed self-energy (\ref{Sigma1(m2)}) is
not well defined and requires regularization in order to be dealt with
in the formalism. According to the method of dimensional regularization,
we start by substituting this ill defined expression in four-dimensional
space-time by the function of $\omega$ given by  
\begin{eqnarray} \label{preAutoEnergiaReg1}
\frac{2}{ig}\Sigma_{1}(m^2,\omega)=(\mu^2)^{2-\omega}\int  \frac{d^{2\omega} p}{(2\pi)^{2\omega}}\frac{1}{q^{-2}\sinh^2(qp^0)-{\bf p}^2-m^2+i\varepsilon}
\end{eqnarray}
or, more explicitly, by the expression
\begin{eqnarray} \label{AutoEnergiaReg1}
\frac{2}{ig}\left.\Sigma_1(m^2,\omega)\right|_{q=0}=
\frac{(4\pi\mu^2)^{2-\omega}}
{16\pi^4\Gamma(\omega-1/2)}\pi^{3/2}
\int_{-\infty}^{\infty} dp^0\int_0^{\infty}dr^2
\frac{(r^2)^{\omega-3/2}}{q^{-2}\sinh^2(qp^0)-r^2-m^2+i\varepsilon} \; ,
\end{eqnarray}
which is certainly well defined for $1/2<\Re\omega<1$, since 
$q^{-2}\sinh^2(qp^0)\ge (p^0)^2$ for any real $p^0$. Therefore,
we can make the change of integration variable 
$q^{-1}\sinh(qp^0)\mapsto p^0 $ in (\ref{AutoEnergiaReg1}) to obtain
\begin{eqnarray} \label{AutoEnergiaReg2}
\frac{2}{ig}\Sigma_{1}(m^2,\omega)=
\frac{(4\pi^2\mu^2)^{2-\omega}}{16\pi^4}
\int_{-\infty}^{\infty} \frac{dp^0}{\sqrt{1+q^2p_0^2}}
\int_0^{\infty}dr^2
\frac{(r^2)^{\omega-3/2}}{(p^0)^2-r^2-m^2+i\varepsilon} \; ,
\end{eqnarray}
We can see in this expression that the change of integration variable has brought the propagator to its non-deformed version, and generated the factor
$1/\sqrt{1+q^2p_0^2}$ in the integrand. This factor behaves asymptotically
as $1/|p^0|$, and obeys the inequality
\begin{eqnarray}\label{majoramentokappa}
\frac{1}{\sqrt{1+q^2p_0^2}}\leq2\kappa \frac{|p_0|+2\kappa}{p^2+(2\kappa)^2}\;,
\end{eqnarray}
for any ${\bf p}$ that we join to $p^0$ in order to form the four-vector
$p$. This inequality shows that the factor $1/\sqrt{1+q^2p_0^2}$ is majorized 
by a fermion-like propagator, a property which shows that it is safe to get 
from this factor a $\;-1$ contribution to the power-counting. As it is clear by now, and will be confirmed in what follows, it is precisely 
this factor which determines the contribution from the $\kappa$-deformation 
to the regularization of the theory. It is quite interesting to note that this fermion-like propagator has an imaginary mass generated by the deformation parameter, just as the Pauli-Villars propagator which provides the natural regularization in the above mentioned theory in the non-commutative $\kappa$-deformed Minkowski space \cite{KosinskiLukierskiMaslanka00}.
Now, we can specify the domain of (\ref{AutoEnergiaReg2}) in the 
complex $\omega$-plane as being the strip $1/2<\Re\omega<3/2$.
The next step to be taken is the
analytic continuation of this expression to a neighbourhood of 
$\omega=2$, which is accomplished by the same method used in the non-deformed
case, namely, by introducing into the integrand of
(\ref{AutoEnergiaReg2}) a factor of $1$ in the form
$(\partial p^0/\partial p^0+\partial r^2/\partial r^2)/2$,
and performing appropriate partial integrations. In this way, we arrive at
\begin{eqnarray}\label{IntegralAposIdentidadeTHooft}
\frac{2}{ig}\Sigma_{1}(m^2,\omega)=
\frac{(4\pi\mu^2)^{2-\omega}}{16\pi^4\Gamma(\omega-1/2)}
\;\frac{\pi^{3/2}}{\omega-3/2}
\left[m^2\int_{-\infty}^{\infty} dp_0 \int_{0}^{\infty} d(r^2) (r^2)^{\omega-3/2} \right. \nonumber \\ \left. \frac{1}{(p^2-m^2+i\varepsilon)^2(1+q^2p_0^2)^{1/2}}-\frac{1}{2}\int_{-\infty}^{\infty} dp_0 \int_{0}^{\infty} d(r^2) \frac{(r^2)^{\omega-3/2}}{(p^2-m^2+i\varepsilon)(1+q^2p_0^2)^{3/2}}\right]\,,
\hspace{-1.5cm} \nonumber \\
\end{eqnarray}
which is well defined in the strip $1/2<\Re\omega<5/2$ punctured at pole $\omega=3/2$. Consequently, we can take $\omega=2$ in (\ref{IntegralAposIdentidadeTHooft}) to obtain a finite result at the physical dimension $2\omega=4$ of space-time. Let us notice that there still is a pole in (\ref{IntegralAposIdentidadeTHooft}), but it is at dimension $2\omega=3$, which is not the physical dimension of
the field in consideration. At the limit of no deformation, $q\rightarrow 0$,
(\ref{IntegralAposIdentidadeTHooft}) reduces to an expression defined
on the strip $1/2<\Re\omega<1$. By using in this expression the result
(\ref{Sigma1(m2,omega)q=0tH-V}), we obtain, after a trivial analytic 
continuation, the non-deformed intermediate result (\ref{Sigma1(m2,omega)q=0prefinal}). In order to obtain the final
non-deformed result (\ref{Sigma1(m2,omega)q=0final}) as a limit of
the deformed self-energy, (\ref{IntegralAposIdentidadeTHooft}) must be
further continued, despite already being finite at $\omega=2$.
Moreover, this further continuation of (\ref{IntegralAposIdentidadeTHooft})  will give us more insight into the regularizing effect of the $\kappa$-deformation.
Following the above described method, we obtain
\begin{eqnarray}\label{IntegralIAposIDTHoft2}
\frac{2}{ig}\Sigma_{1}(m^2,\omega)=
\frac{(4\pi\mu^2)^{2-\omega}}{16\pi^4\Gamma(\omega-1/2)}
\;\frac{\pi^{3/2}}{\left(\omega-3/2\right)\left(\omega-5/2\right)}
\left[2m^4\int_{-\infty}^{\infty} dp_0 \int_{0}^{\infty} d(r^2) \hspace{1cm} \right. \nonumber \\
\left. \frac{(r^2)^{\omega-3/2}}{(p^2-m^2+i\varepsilon)^{3}(1+q^2p_0^2)^{1/2}}
-m^2\int_{-\infty}^{\infty} dp_0 \int_{0}^{\infty} d(r^2)\frac{(r^2)^{\omega-3/2}}{(p^2-m^2+i\varepsilon)^{2}
(1+q^2p_0^2)^{3/2}} \right. \nonumber \\
\left. +\frac{3}{4}\int_{-\infty}^{\infty} dp_0 \int_{0}^{\infty} d(r^2)\frac{(r^2)^{\omega-3/2}}{(p^2-m^2+i\varepsilon)(1+q^2p_0^2)^{5/2}}\right] , \hspace{0.5cm}
\end{eqnarray}
which is well defined in the strip $1/2<\Re\omega<7/2$ punctured at the poles
$\omega=3/2$ and $\omega=5/2$. Taking the limit of this function when
$q\rightarrow 0$, we obtain again an expression defined on the strip $1/2<\Re\omega<1$. By using in this expression the results
(\ref{Sigma1(m2,omega)q=0tH-V}) and (\ref{Sigma1(m2,omega)q=0prefinal}), we obtain, after a trivial analytic continuation, the non-deformed final result (\ref{Sigma1(m2,omega)q=0final}). The comparison of the non-deformed 
result (\ref{Sigma1(m2,omega)q=0final}) with the deformed one
(\ref{IntegralIAposIDTHoft2}) shows that the effect of the deformation on the analytic continuation (\ref{Sigma1(m2,omega)q=0final}) is to widen its domain,
and shift its poles $1/2$ to the right. The $\kappa$-deformation shifts the
pole at $\omega=1$ to $\omega=3/2$, and the pole $\omega=2$ corresponding
to physical dimension to $\omega=5/2$. The calculations already done also
show that this shift of poles to the right is due to the deformation 
factor $1/\sqrt{1+q^2p_0^2}$ which is majored by the fermion-like propagator according to (\ref{majoramentokappa}). It is fairly safe to guess that this 
shift of $1/2$ to the right will occur to all poles by effect of the deformation. Due to this shift, we obtain in the limit of physical dimension $2\omega\rightarrow 4$ that the analytic continuations 
(\ref{IntegralAposIdentidadeTHooft}) or (\ref{IntegralIAposIDTHoft2}) have
the finite value
\begin{eqnarray}\label{AutoEnergiaAposIdTHooftomega2}
\frac{2}{ig}\Sigma_{1}(m^2)=-\frac{1}{2\pi^4}\left[ 2m^4\int_{-\infty}^{\infty} dp_0 \int_{0}^{\infty} \! d(r^2) \frac{r}{(p^2-m^2+i\varepsilon)^{3}(1+q^2p_0^2)^{1/2}} \right. \nonumber \\
\left.-\,m^2\int_{-\infty}^{\infty} dp_0 \int_{0}^{\infty} \! d(r^2)\frac{r}{(p^2-m^2+i\varepsilon)^{2}(1+q^2p_0^2)^{3/2}} \right. \nonumber \\
\left. +\,\frac{3}{4}\int_{-\infty}^{\infty} dp_0 \int_{0}^{\infty} d(r^2)\frac{r}{(p^2-m^2+i\varepsilon)(1+q^2p_0^2)^{5/2}}\right] \; . \hspace{0.5cm}
\end{eqnarray}
The limit of this quantity when $q\rightarrow 0$ is the usual divergent
self-energy in the non-deformed case.

Now, let us consider the four-point function (\ref{FourierDelta4}), which
is not well defined due to its dependence on the three divergent vertex integrals ${\Gamma}^{(4)}(m,s,\omega)$, ${\Gamma}^{(4)}(m,t,\omega)$ and
${\Gamma}^{(4)}(m,u,\omega)$ defined by
\begin{eqnarray}\label{Gamma4(m2,s,omega}
\frac{2}{g^2}{\Gamma}^{(4)}(m,s,\omega)=
\int \frac{d^4p}{(2\pi)^{4}}\frac{1}{\left[(p-s)^{\bar{\mu}}(p-s)_{\bar{\mu}}-m^2+i\varepsilon\right](p^{\bar{\mu}}p_{\bar{\mu}}-m^2+i\varepsilon)} \; ,
\end{eqnarray}
and the expressions obtained by replacing $s$ by $t$ or $u$. 
According
to the method of dimensional regularization, the expression 
(\ref{Gamma4(m2,s,omega}) should be substituted by the function of
$\omega$
\begin{eqnarray}\label{FuncaoVertice4PtosSegundaOrdemReg}
\frac{2}{g^2}{\Gamma}^{(4)}(m,s,\omega)=(\mu^2)^{2-\omega}\!\!\int \frac{d^{2\omega}p}{(2\pi)^{2\omega}}\frac{1}{[(p-s)^{\bar{\mu}}(p-s)_{\bar{\mu}}-m^2+i\varepsilon](p_{\bar{\mu}}p^{\bar{\mu}}-m^2+i\varepsilon)} \; , 
\hspace{-2cm} \nonumber \\
\end{eqnarray}
which is certainly well defined for $1/2<\Re\omega<2$. By making the change of integration variable $q^{-1}\sinh(qp^0) \mapsto p^0$ we obtain from 
(\ref{FuncaoVertice4PtosSegundaOrdemReg}) the expression
\begin{eqnarray}\label{FuncaoVertice4PtosSegundaOrdemRegTransformp0}
\frac{2}{g^2}{\Gamma}^{(4)}(m,s,\omega)=(\mu^2)^{2-\omega}\!\!\int \frac{d^{2\omega}p}{(2\pi)^{2\omega}}\frac{1}{\sqrt{1+q^2p^2_0}}
\frac{1}{p^2-m^2+i\varepsilon} \times \hspace{1.5cm}  \nonumber \\
\frac{1}{p_0^2-q^{-1}\sinh(2qs^0)p_0\sqrt{1+q^2p_0^2}+(1+2q^2p_0^2)
q^{-2}\sinh^2(qs^0)-({\bf p}-{\bf s})^2-m^2+i\varepsilon}\; , \hspace{0.5cm}
\end{eqnarray}
which has a trivial analytic continuation from $1/2<\Re\omega<2$ to $1/2<\Re\omega<5/2$. Since  (\ref{FuncaoVertice4PtosSegundaOrdemRegTransformp0}) continued to this wider domain is regular at $\omega=2$, we immediately obtain  for the physical dimension the finite quantity
\begin{eqnarray}\label{FuncaoVertice4PtosSegundaOrdemRegTransformp0F}
\frac{2}{g^2}{\Gamma}^{(4)}(m,s,2)=\int \frac{d^{4}p}{(2\pi)^4}\frac{1}{\sqrt{1+q^2p^2_0}}\frac{1}{p^2-m^2+i\varepsilon} \times \hspace{1.5cm}  \nonumber \\
\frac{1}{p_0^2-q^{-1}\sinh(2qs^0)p_0\sqrt{1+q^2p_0^2}+(1+2q^2p_0^2)
q^{-2}\sinh^2(qs^0)-({\bf p}-{\bf s})^2-m^2+i\varepsilon}\;. \hspace{0.5cm}
\end{eqnarray}
It should come as no surprise that this result is obtained 
from trivial analytic continuation with no need of partial integrations.
In fact, since (\ref{Gamma4(m2,s,omega}) with its logarithmic divergence is at
the threshold of convergence, the deformation alone is capable of bringing it
to convergence at physical dimension $2\omega=4$. 
From our previous 
experience with the two-point function, we expect that the pole at $\omega=2$
of the non-deformed analytic continuation  
$\left.{\Gamma}^{(4)}(m,s,\omega)\right|_{q=0}$ 
is shifted by the deformation to $\omega=5/2$. 

\section{Conclusion}

We have considered a $\kappa$-deformed scalar field with quartic
self-interaction to determine a possible regularizing effect of
the deformation on the otherwise primitively divergent
diagrams of the theory. Although the deformation by itself is manifestly not
sufficient to render the diagrams finite, it was found that the
deformation leads to finite diagrams at physical dimension after 
the usual analytic continuations are performed on dimensionally 
regularized diagrams. 

It would be interesting to check whether finite 
results are obtained by using other regularization methods; we were 
unable to calculate all the required diagrams with other methods, 
although the simplest ones that could be calculated turned out to be finite.

Nevertheless, the association between $\kappa$-deformation and dimensional regularization that leads to finite diagrams seems natural, since both deformation and regularization affect the properties of space-time. Indeed, dimensional regularization 
extends space-time dimension to the complex plane, while $\kappa$-deformation
of the symmetry algebra of space-time has the effect of shifting poles
in this plane, as we saw in our calculations.

It is an elementary fact that regularization procedures can lead to finite results. Especially interesting are the examples provided by analytic 
regularization methods (for a review see, {\it e.g.}, \cite{Elizalde94,Elizalde03}). 
For example, the sum $\sum_{n=1}^\infty n^{-1}$ can be 
regularized as $\sum_{n=1}^\infty n^{-1+\omega}$ with $\Re\omega<0$,
and analytically continued to a neighbourhood of $\omega=0$, where 
we find the divergent result $\zeta(1)=\infty$, while 
the sum $\sum_{n=1}^\infty n$ can be regularized as
$\sum_{n=1}^\infty n^{1+\omega}$ with $\Re\omega<-2$,
and analytically continued to a neighbourhood of $\omega=0$, where 
we find the finite result $\zeta(-1)=-1/12$, where $\zeta$ is the usual
Riemann zeta function. 

In the same way we may say 
that dimensional regularization applied to non-deformed diagrams leads to infinities which require subtractions, while it leads to finite expressions 
when applied to $\kappa$-deformed diagrams. What is interesting in the
$\kappa$-deformed theory is that the finite expressions result from
a general theory deforming space-time symmetries, and the mechanism
by which the deformation leads to finite results is simple and clear,
to wit: the poles in the complex plane of dimensions are shifted
by $1/2$, in particular from the physical dimension at $\omega=2$ to
$\omega=5/2$. 

Although those mathematical considerations are instructive, it
would be most interesting to understand the physical reason behind
the finite expressions of the $\kappa$-deformed theory. A possible reason
has been advanced in the early works on $\kappa$-deformed Poincaré algebra
\cite{LukierskiNowickiRuegg92,LukierskiRueggZakrewski95}.
There was pointed out that the $q$-differential operator $\partial_q$ which is
defined in (\ref{partialq}), and present in the $\kappa$-deformed Klein-Gordon equation (\ref{pospreqKleinGordon}), generates the time translation in finite jumps. Actually, it acts on a function $f$ of time as a finite difference
operator with symmetrical shifts of size $q$ along imaginary time,
$\partial_qf(t)=[f(t+iq)-f(t-iq)]/(2q)$. Sure enough, this is not
a discretization of time, with its obvious effect in softening the
divergences of the theory, but a discretization of time 
evolution -- that could be a physical reason for the elimination of 
divergences in a $\kappa$-deformed theory dimensionally regularized. 
We may also consider situations in which the $\kappa$-deformation 
appears as a factor of suppression of vacuum fluctuations and, as such,
as a physical reason for softening of divergences. We have seen
that the deformation parameter $\kappa$ gives rise to Pauli-Villars
masses, as in \cite{KosinskiLukierskiMaslanka00} and in the fermion-like propagator mentioned after inequality (\ref{majoramentokappa}).
The effective action of a $\kappa$-deformed scalar field under Dirichlet boundary conditions 
on parallel planes \cite{{C-PFarina97}} gives rise to a real part
proportional to the Casimir energy \cite{Casimir48} (for reviews see, {\it e.g.},
\cite{BordagMohideenMostepanenko01,Milton01-Milton04}) and an 
imaginary part proportional to a creation rate of field excitations.
The resulting Casimir energy, which may be viewed as a measure of vacuum
fluctuations, is exponentially damped by the squared mass of the field,
as is usual in the Casimir effect. However, due to the $\kappa$-deformation,
a term $2\kappa^2$ is added to the mass, thereby enhancing the damping of
the Casimir energy. 
A damping is also observed in the calculation of the Casimir energy
of a $\kappa$-deformed electromagnetic field \cite{C-PFarinaMendes02-C-PFarinaMendes04}.
At any rate, the elimination of divergences after dimensional regularization
presented here is an interesting feature of a theory with a fundamental length, which shows that the introduction of such a length can simplify at least some aspects of the theory. 

\end{document}